\documentclass[epj]{svjour}
\usepackage{graphics,amsmath,enumitem,array,multirow,bm,ulem,enumitem,slashed,xcolor,setspace,booktabs,hyperref,cite,url,tabularx, float,tikz-feynman}
\usepackage{subcaption}
\usepackage{setspace}
\graphicspath{ {Figures/} }
\begin{document}
\title{
%Tri-Boson Excesses in Light of Drell-Yan Production of Triplet Higgses
Multi-Lepton Probes of the Drell-Yan Production of Triplet Higgses}

\author{
Siddharth P. Maharathy
\inst{1,2,}\thanks{\email{siddharth.prasad.maharathy@cern.ch}} 
\and 
Srimoy Bhattacharya
\inst{1,}\thanks{\email{srimoy.bhattacharya@cern.ch}}
\and 
Andreas Crivellin
\inst{3,4,}\thanks{\email{andreas.crivellin@cern.ch}}
\and 
Mukesh Kumar
\inst{1,}\thanks{\email{mukesh.kumar@cern.ch}}
\and 
Rachid Mazini
\inst{1,}\thanks{\email{rachid.mazini@cern.ch}}
\and 
Bruce Mellado
\inst{5,1,}\thanks{\email{bmellado@mail.cern.ch}}}                     

\institute{School of Physics and Institute for Collider Particle Physics, University of the Witwatersrand, Johannesburg, Wits 2050, South Africa  
%\and iThemba LABS, National Research Foundation, PO Box 722, Somerset West 7129, South Africa
\and Indian Institute of Science Education and Research Pune, Dr.~Homi Bhabha Road, Pune 411008, India
\and Universitat Autònoma de Barcelona, 08193 Bellaterra, Barcelona
\and ICREA, Instituci\'o Catalana de Recerca i Estudis Avan\c{c}ats, Passeig de Llu\'{\i}s Companys 23, 08010 Barcelona, Spain
\and Institute of High Energy Physics, 19B, Yuquan Road, Shijing District, 100049, Beijing, China}

\abstract{
Excesses in di-photon, $Z\gamma$, and $WW$ spectra indicate the existence of a new Higgs boson with mass $152\pm1$\,GeV. However, no excess is observed in the $ZZ$ channel. This pattern aligns with a Real Higgs Triplet model with hypercharge $Y = 0$ ($\Delta$SM). A prediction of this model is the Drell–Yan production of scalars at the LHC, which dominantly decay to electroweak bosons, thus enhancing the cross sections of triboson channels such as $WWZ$, $WZZ$, and $WWW$. Interestingly, both ATLAS and CMS have reported higher-than-expected significances for such processes:  $6.4\sigma$ (observed) vs $4.7\sigma$ (expected) in the $VVZ$ (where $V = W$ or $Z$) channel and $4.4\sigma$ vs $3.6\sigma$ in $WWZ$, suggesting the possibility that these signals may be the manifestations of an extended Higgs sector. We investigate whether the $\Delta$SM can account for these triboson excesses through electroweak production and decay of triplet scalars. We find that while current data prefers a non-zero new physics signal ($2.6\sigma$), the $\Delta$SM predicts more events than observed, such that it is consistent with data but not preferred over the SM. However, this
tension could be clarified with Run~3 and HL-LHC data.} 
\date{}
\maketitle
 \section{Introduction}\label{sec:intro}

The 2012 discovery of the Higgs boson~\cite{ATLAS:2012yve,CMS:2012qbp} was a landmark achievement that completed the particle content of the Standard Model (SM). Despite its success, growing experimental anomalies suggest that the SM is not the complete description of nature~\cite{Crivellin:2023zui}. The so-called ``Multilepton anomalies" — statistically significant excesses observed in LHC data across several channels (see Refs.~\cite{Fischer:2021sqw,Crivellin:2023zui} for recent reviews). These events feature multiple leptons (electrons and/or muons), moderate missing transverse energy, and ($b$) jets. The consistency of these anomalies across different final states points toward the presence of additional scalar resonances~\cite{vonBuddenbrock:2016rmr,vonBuddenbrock:2017gvy,Hernandez:2019geu,Buddenbrock:2019tua}. In particular, the mass of one of the scalars is predicted to be $m_S=150\pm5$\,GeV~\cite{vonBuddenbrock:2017gvy}. In this context, recent measurements have reported excesses in di-photon~\cite{ATLAS:2021jbf,ATLAS:2023omk}, $Z\gamma$~\cite{CMS:2018myz,CMS:2022ahq}, and $WW$~\cite{CMS:2022uhn,ATLAS:2022ooq,Coloretti:2023wng} final states, providing strong indications in support of a new scalar particle with a mass $152\pm1$\,GeV~\cite{Crivellin:2021ubm,Bhattacharya:2025rfr}. Interestingly, no corresponding excess has been seen in the $ZZ$ channel~\cite{CMS:2022dwd, ATLAS:2020rej}, a pattern that poses a challenge for models involving a standard Higgs doublet or singlet extension. However, this pattern is naturally obtained if the new scalar is the neutral component of a hypercharge $Y = 0$ triplet~\cite{Ross:1975fq,Gunion:1989ci} and, in fact, the observed excesses can be partially explained by the Drell-Yan production of the triplet~\cite{Ashanujjaman:2024pky, Crivellin:2024uhc,Ashanujjaman:2024lnr,Maharathy:2025yyo}. 

\begin{figure}[htb!]
\centering
 \resizebox{0.45\textwidth}{!}{\includegraphics{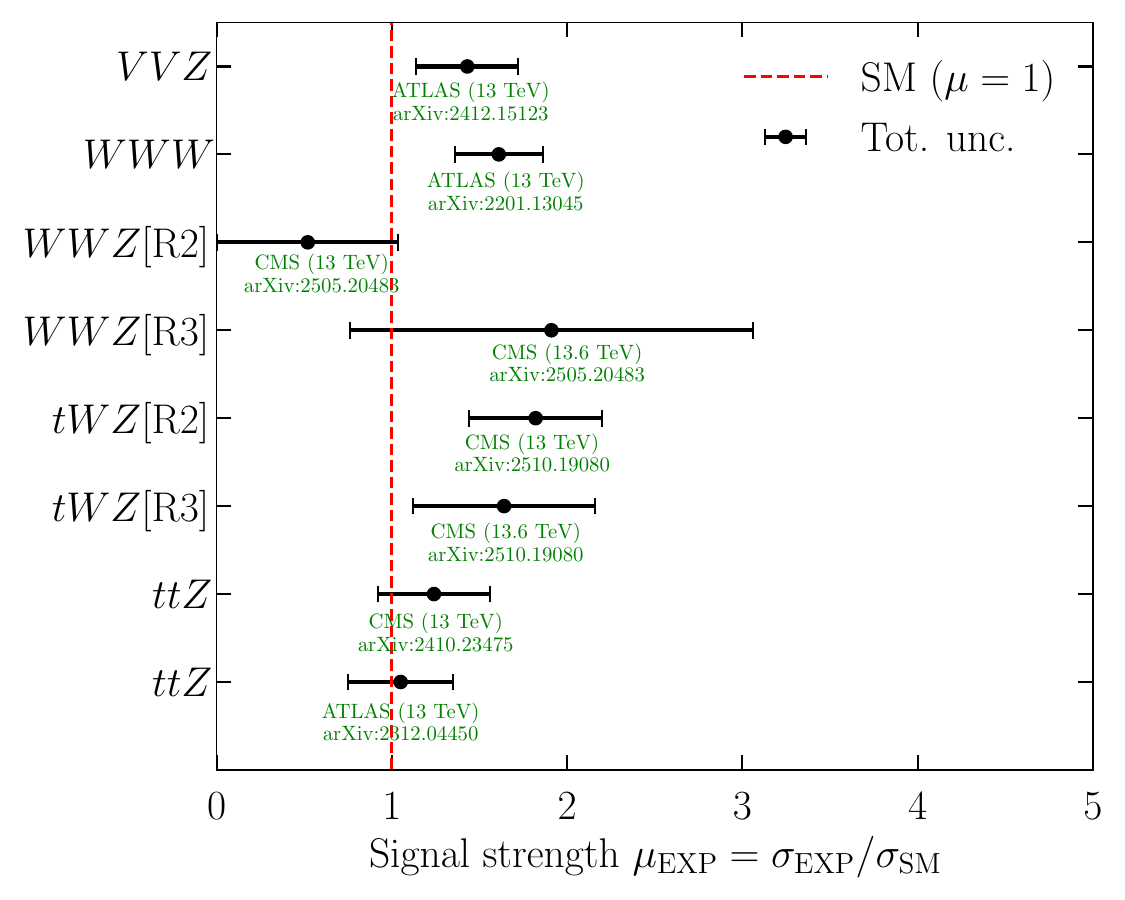}}
\caption{Measured signal strengths relative to the expected SM values ($\mu_{\rm EXP} = \sigma_{\rm EXP}/\sigma_{\rm SM}$) for each analysis along with their uncertainties. The total uncertainty is obtained by adding the statistical, systematic and the theory error in quadrature. R2  and R3 stand for  Run~2 and Run~3, respectively. It is interesting to note that except for the Run 2 CMS analysis of $WWZ$, all values are above 1.
}
\label{fig:meassured_signal_strength}
\end{figure}

The Drell-Yan production of the neutral and charged components of the triplet can not only explain the excesses in associated di-photon production, but via its dominant decays to electroweak (EW) gauge bosons also leads to $WWW$, $WWZ$, and $WZZ$ final states, which are currently under close experimental scrutiny. Intriguingly, both ATLAS and CMS have reported higher-than-expected significances for these processes. For example, $6.4\sigma$ vs $4.7\sigma$ in the $VVZ$ channel~\cite{ATLAS:2024nab} and $4.4\sigma$ vs $(3.6\sigma)$ in $WWZ$~\cite{CMS:2025hlu}. {Since the experimental searches are statistically uncorrelated, and the statistical uncertainties dominate over the systematic ones, any correlations among them are expected to be small. The ``multi-lepton anomalies” suggest the existence of a new scalar with a mass range of $150 \pm 5$ GeV that decays dominantly to $W$ bosons and is produced in association with lepton, bottom quarks, and missing energy. Both the production and decay modes are in agreement with the triplet hypothesis. Importantly,
these indirect hints for a new scalar reduce the look-else where effect.} In this analysis, we have not considered the CMS $WWW$ analysis, since the CMS analysis requires a large $\Delta \phi$ cut requiring events with higher $WWW$ invariant mass, which is not sensitive to NP at lower masses. This is illustrated in Table~\ref{tab:xsections_sensitivity} and Fig.~\ref{fig:meassured_signal_strength} where the measured signal strengths for these processes w.r.t.~their values expected in the SM are shown.

So far, it has been shown that the SM supplemented with a $Y=0$ triplet, the $\Delta$SM is not excluded from multi-lepton searches~\cite{Ashanujjaman:2024lnr,Butterworth:2023rnw}. In this paper, we investigate whether the $\Delta$SM, can accommodate the elevated multi-boson cross sections. We focus on the parameter space where the neutral component has a mass $152\pm1$\,GeV, as suggested by the di-photon and $Z\gamma$ measurements, and the charged component must be quasi mass degenerate because of EW precision constraints. Using detailed Monte Carlo simulations, we analyse the production and decay of these triplet scalars and compute the predicted cross-sections for $WWZ$, $WZZ$, $WWW$ and $tWZ$ final states. We compare these predictions with the latest LHC measurements to assess the viability of the $\Delta$SM in this context.

\section{Signal regions and validation}

\begin{table*}[t!]
\centering
% \resizebox{1.\textwidth}{!}
{
\begin{tabular}{c|c|c|c|c}
\toprule \toprule
CoM Energy & Process  & Measured Cross Section [fb] &  SM Cross Section [fb]  & $\mu_{\rm EXP} = \sigma_{\rm EXP}/\sigma_{\rm SM}$\\\hline
 $\sqrt{s}=13~\text{TeV}$ & \rule{0pt}{14pt} $VVZ$[ATLAS] & $660^{+93}_{-90}(\text{stat.})^{+88}_{-81}(\text{syst.})$~\cite{ATLAS:2024nab} &  $461 \pm 23$~\cite{ATLAS:2024nab}  & $1.43\pm 0.29$\\
 \cline{2-5} &\rule{0pt}{14pt} $WWW$[ATLAS]  & $820 \pm 100(\text{stat.})\pm 80(\text{syst.})$~\cite{ATLAS:2022xnu}  & 511.0 $\pm$ 18~\cite{ATLAS:2022xnu} & $1.61\pm 0.25$\\
  \cline{2-5} &\rule{0pt}{14pt} $WWZ$[CMS]  & $97^{+91}_{-75}(\text{stat.})^{+24}_{-21}(\text{syst.}) $~\cite{CMS:2025hlu}  & $184 \pm 9.2$~\cite{CMS:2025hlu} & $0.52\pm 0.52$\\
 \cline{2-5}
&\rule{0pt}{14pt} $tWZ$[CMS]  & $248 \pm 38(\text{stat.})\pm 35(\text{syst.})$~\cite{CMS:2025xph}  & $136.0^{+9}_{-8}$~\cite{CMS:2025xph} & $1.82\pm 0.38$\\
 \hline
  \cline{2-5}
&\rule{0pt}{14pt} $ttZ$ [CMS]  & $1224.6 \pm 156.7$~\cite{CMS:2024mke}  & $989 \pm 49 $~\cite{CMS:2024mke} & $1.24 \pm 0.32$\\
 \hline
   \cline{2-5}
&\rule{0pt}{14pt} $ttZ$ [ATLAS]  & $860 \pm 40\text{(stat.)} \pm 40\text{(syst.)} $~\cite{ATLAS:2023eld}  & $860^{+80}_{-90} $~\cite{ATLAS:2023eld} & $1.05 \pm 0.30$\\
 \hline
$\sqrt{s}=13.6~\text{TeV}$  & $WWZ$ [CMS]& $380^{+220}_{-190}\text{(stat)}^{+50}_{-30}\text{(syst)}$~\cite{CMS:2025hlu} & $200 \pm 10$  \cite{CMS:2025hlu} & $1.91\pm 1.15$\\
\cline{2-5}
  & $tWZ$ [CMS]& $242 \pm 62(\text{stat.})\pm 46(\text{syst.})$~\cite{CMS:2025xph} & $147.8^{+10}_{-9}$~\cite{CMS:2025xph} & $1.64\pm 0.52$\\
\bottomrule \bottomrule
\end{tabular}
}
\caption{Observed and expected SM values for the inclusive cross sections of $VVZ$, $WWW$, $tWZ$ and $ttZ$ production at the LHC. The total uncertainty of $\mu_{\rm EXP}$ is obtained by adding all errors in quadrature.}
\label{tab:xsections_sensitivity}
\end{table*}

In this article, we study analyses targeting three EW gauge bosons, $V=W,Z$, produced directly or indirectly via top decays done in the SM context. In Table~\ref{tab:xsections_sensitivity}, we summarise the ATLAS and CMS results for the different analyses that we consider in this work. Furthermore, the measured signal strengths normalised to their SM predictions ($\mu_{\rm EXP}$) for each process is illustrated in Fig~\ref{fig:meassured_signal_strength}.\footnote{For the $ttZ$ signal strength, we used the digitized data of Refs.~\cite{CMS:2024mke, ATLAS:2023eld}. } 

To validate the setup for our analysis, we first simulated the SM processes for the different analyses using {\tt MadGraph5\_aMC\_v3.5.3}~\cite{Alwall:2014hca,Frederix:2018nkq} for event generation. The obtained parton-level events are passed through the regular chain of tools, namely {\tt Pythia 8.3}~\cite{Sjostrand:2014zea} and {\tt Delphes 3.5.0}~\cite{deFavereau:2013fsa} (we will thus abbreviate our simulation with MPD), to obtain the result of subsequent decays of the unstable particles, radiations, showering, fragmentation, hadronisation as well as various detector effects and particle-level object reconstruction. We then implemented the respective cuts for each analysis. We explain each of the considered analyses in detail below.

The cutflows for the different signal regions for each considered analysis are shown in the appendix in Tables~\ref{tab:ATLASVVZ_3l}--\ref{tab:CMStWZ_R3}.

\subsection{ATLAS $(VVZ)$}

The ATLAS $VVZ$ analysis consists of three signal regions (SRs)~\cite{ATLAS:2024nab} containing three ($3\ell$), four ($4\ell$) and five ($5\ell$) charged leptons ($\ell=e,\mu$) in the final state. These leptons are required to have a rapidity of $|\eta|<2.47$ for electrons and $|\eta|<2.7$ for muons. The transverse-momentum threshold for the leptons in the $3\ell$ SR are $p_T>27$\,GeV, $15$\,GeV, and $15$\,GeV. Similarly, for the $4\ell$ SR, ATLAS requires 4-leading leptons with $p_T>30$\,GeV, $15$\,GeV, $8$\,GeV, and $6$\,GeV. A $Z$-boson candidate is identified from a same-flavor opposite-sign lepton pair with an invariant mass requirement of $|m_{\ell\ell} - m_Z| < 20$\,GeV, where $m_{\ell\ell}$ is the reconstructed same-flavor opposite-sign di-lepton invariant mass. Jets are reconstructed using the anti-$k_T$ algorithm with radius parameter $R=0.4$, requiring $p_T^j>20$\,GeV and $|\eta_j|<5.0$. The overlap removal of leptons and jets is implemented according to the description of Ref.~\cite{ATLAS:2024nab}. For the $3\ell$ SR, in addition to the three leptons, at least one light jet is required, while $b$-tagged jets are vetoed in all three SRs.

To validate our setup, we simulated the SM processes $pp \to VVZ$, where $V = W, Z$. The number of expected events from the ATLAS simulation and our MPD one for the $3\ell, 4\ell$ and $5\ell$ signal regions for an integrated luminosity of 139 fb$^{-1}$ and 13\,TeV center of mass (CoM) are
\begin{equation}
 \begin{tabular}{c|c|cc}
$VVZ$ SM & $3\ell$  & $4\ell$ &  $5\ell$  \\
\hline
ATLAS~\cite{ATLAS:2024nab}& 376 & 72 & 4 \\
MPD & 366 & 55 & 3 \\
\end{tabular}
\end{equation}
Summing the events in all signal regions, we find
\begin{equation}
\frac{N_{\rm SM}^{\rm ATLAS}}{N_{\rm SM}^{\rm MPD}} \approx  1.07\,.
\end{equation} 
The proximity to unity of this correction factor shows good agreement of our fast simulation with the full detector simulation of ATLAS. Since our analysis relies on the overall signal strength from the ATLAS $VVZ$ measurement (see Table~\ref{tab:xsections_sensitivity}), and the individual signal strengths for each signal region are not provided, summing over all signal regions offers a reasonable approximation for comparison with our simulation. 

\subsection{ATLAS $(WWW)$}

The signal regions considered in this analysis require events with exactly two and three leptons~\cite{ATLAS:2022xnu}. Electrons (muons) are required to satisfy $p_T>20$\,GeV and $|\eta|<2.47$ $(|\eta|<2.5)$ excluding 
electrons within $1.37 < |\eta|< 1.52$. Jets are reconstructed using the anti-$k_T$ algorithm with $R=0.4$, requiring $p_T^j>30$\,GeV in the forward direction ($2.5<|\eta_j|<4.5$) and $p_T^j>20$\,GeV in the central region ($(|\eta|<2.5)$). For the $2\ell$ SR events are required to have exactly two same-charged leptons and at least two central jets with no $b$-tagged jets. The transverse-momentum of the leading lepton and the dilepton invariant mass should have $p_T > 27$\,GeV and $40\,\text{GeV} < m_{\ell\ell} < 400\,\text{GeV}$. The required condition on the dijet invariant mass and the pseudorapidity separation between the two jets is $m_{jj} < 160\,\text{GeV}$ and $|\Delta\eta_{jj}| < 1.5$. Depending on the flavor of the two leptons, the $2\ell$ SR is further sub-categorized into: $ee$-mode, $e\mu$-mode, and $\mu\mu$-mode. The $3\ell$ SR events are selected requiring exactly three leptons with the leading lepton having $p_T  > 27$\,GeV, no $b-$tagged jets, and no same-flavor opposite-sign (SFOS) lepton pair. 

For the SM $WWW$ simulation, we follow the same procedure as we followed for the SM $VVZ$ process i.e, the montecarlo events $pp \to WWW$ was generated by Madgraph, hadronization is performed by Pythia and detector response is considered by Delphes.  The total number of events in the signal regions for the $WWW$ process from the ATLAS analysis and our simulation compared to ATLAS is 
\begin{equation}
    \begin{tabular}{c|c|cc}
$WWW$ SM & $2\ell$  & $3\ell$   \\
\hline
ATLAS~\cite{ATLAS:2022xnu}& 234 & 35 \\
MPD & 230 & 38 \\
\end{tabular}
\end{equation}
which shows a good agreement
\begin{equation}
\frac{N_{\rm SM}^{\rm ATLAS}}{N_{\rm SM}^{\rm MPD}} \approx 1.004\,.
\end{equation}

To conclude this section, it is important to note that the corresponding results reported by the CMS experiment in Ref.~\cite{CMS:2020hjs} are not included in this study. This is motivated by the presence of selection requirements, such as those on the momentum of the three-lepton system and on the azimuthal angle between this system and the missing transverse momentum, which suppress the efficiency for the new physics scenarios considered here.

\subsection{CMS ($WWZ$)}

The CMS cross-section measurements of $WWZ$~\cite{CMS:2025hlu} production was done at $\sqrt{s} = 13$\,TeV and $13.6$\,TeV, corresponding to a total integrated luminosity of $200$ fb$^{-1}$. The main physics objects used in the analysis are leptons ($e$ and $\mu$), escaping particles ($p_T^{\rm miss}$), and jets. The final states contain two $W$ bosons and a $Z$ boson decaying to four isolated leptons, either directly or via intermediate tau decays. The leading (sub-leading) lepton in the event is required to have $p_T > 25 (15)$\,GeV, and the $p_T$ of the third and fourth lepton is required to have  $p_T > 10$\,GeV. Muons and electrons are required to have $|\eta| < 2.4$ and $2.5$, respectively. Two of the four leptons are demanded to be consistent with originating from the decay of an on-shell $Z$ boson: These two leptons, referred to as ``$Z$ candidate leptons", are required to be of the same flavor, to have opposite charges, and to form an invariant mass within $10$\,GeV of the known $Z$ boson mass. The remaining two leptons must have opposite charges, and are referred to as ``$W$ candidate leptons". Depending upon the flavours of the $W$ candidate leptons, events are further categorized as ``opposite flavor" (OF: $e\mu$) or ``same flavor" (SF: $ee$ or $\mu\mu$). In the SF channel, the $W$ candidate leptons are required to form an invariant mass more than $10$ GeV away from the known mass of the $Z$ boson; additionally, the $m_{T2}$ variable \cite{Lester:1999tx, Barr:2003rg} (which is constructed from the transverse momenta of the $W$ candidate leptons and the $p_T^{\rm miss}$ ) must be greater than $25$\,GeV. Following the event selection requirement, the simulated SM
\begin{equation}
\begin{tabular}{c|c|cc}
$WWZ$ SM & Run~2 & Run~3   \\
& [138 $\text{fb}^{-1}$]  & [62 $\text{fb}^{-1}$]   \\
\hline
CMS [OF]& 7.29 & 3.63 \\
CMS [SF]& 4.38 & 2.2 \\
MPD [OF]&6.1 & 2.4 \\
MPD [SF]& 3.4 & 1.3 \\
\end{tabular}
\end{equation}
For Run~2 and Run~3 simulations we have considered $\sqrt{s} = 13$\,TeV and $13.6$\,TeV with integrated luminosities $138\,{\rm fb}^{-1}$ and $62\,{\rm fb}^{-1}$, respectively. The ratios for the number of events from the full detector simulations of ATLAS vs our fast simulation are 
\begin{equation}
\begin{aligned}
\frac{N_{\rm SM}^{\rm ATLAS}}{N_{\rm SM}^{\rm MPD}} &\approx  1.4\; ({\rm R2})\\
\frac{N_{\rm SM}^{\rm ATLAS}}{N_{\rm SM}^{\rm MPD}} &\approx  1.6 \;({\rm R3})
\end{aligned}
\end{equation}
Taking into account that here 4 leptons are required, which makes the analysis very sensitive to efficiency and acceptance effects, the agreement is reasonable.

\begin{figure*}[t!]
\centering
 \resizebox{0.45\textwidth}{!}{\includegraphics{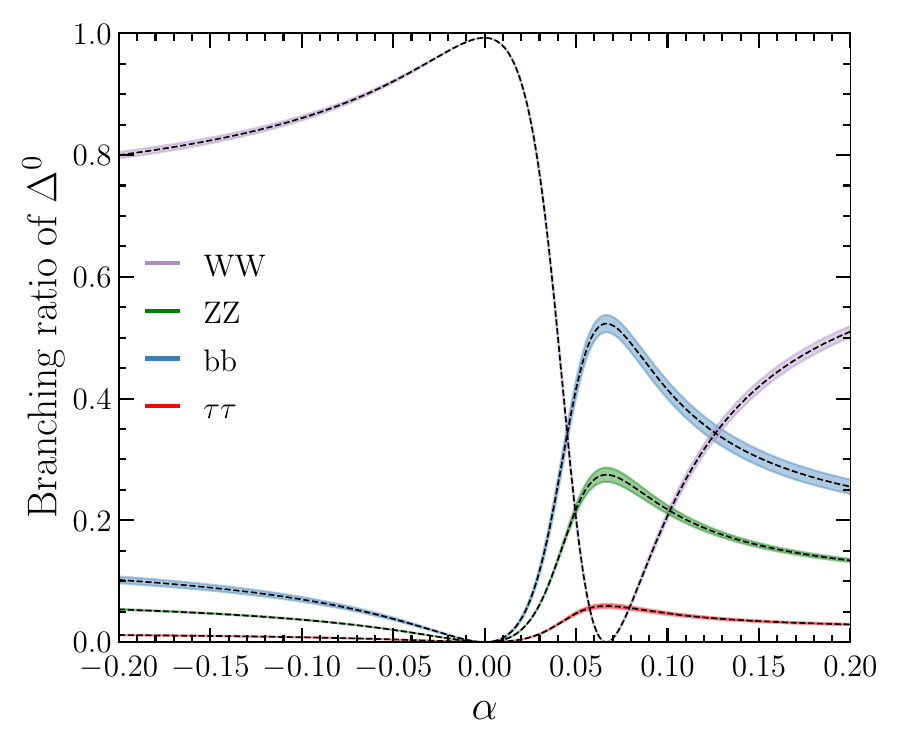}}    
  \resizebox{0.45\textwidth}{!}{\includegraphics{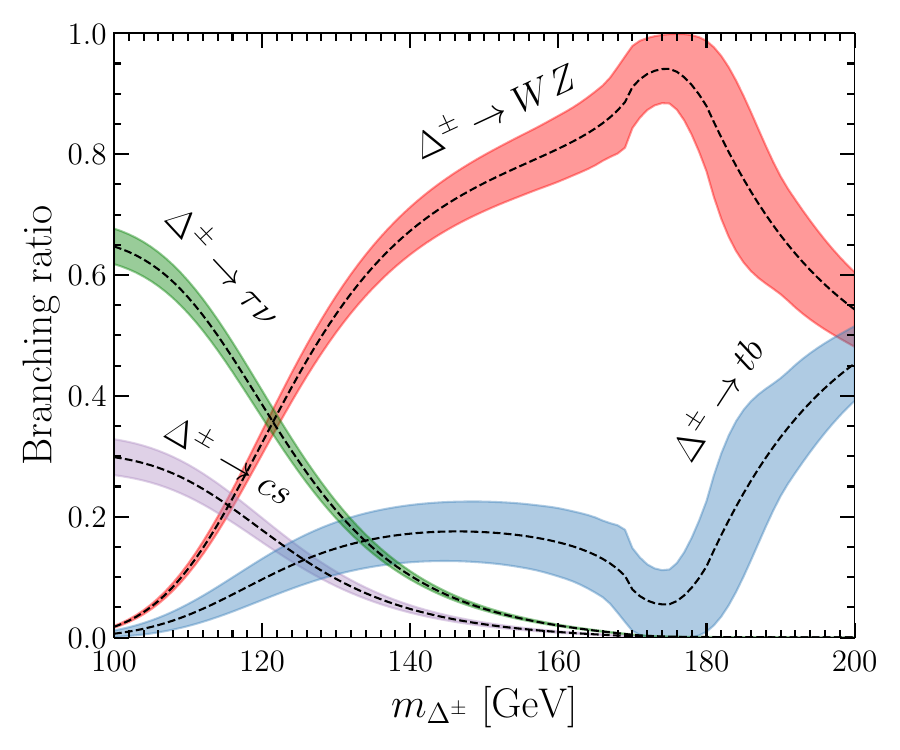}}  
\caption{Left: Dominant branching ratios of $\Delta^0$, including the uncertainties estimated from Ref.~\cite{LHCHiggsCrossSectionWorkingGroup:2013rie}, and $m_{\Delta^0} = 150$\,GeV. Right: Dominant branching ratios of $\Delta^\pm$, including the uncertainties estimated from Ref.~\cite{LHCHiggsCrossSectionWorkingGroup:2013rie}, as a function of its mass. We have assumed $\alpha \approx 0$.}
\label{fig:Br_150_neutral}
\end{figure*}

\subsection{CMS ($tWZ$)}

The CMS analysis~\cite{CMS:2025xph} provides the first observation of single top quark production in association with a $W$ and a $Z$ boson. The data used for this analysis corresponds to center-of-mass energies of $13$\,TeV and $13.6$\,TeV with a total integrated luminosity of 200 ${\rm fb}^{-1}$. The leptons ($\ell = e, \mu$) are required to have transverse momentum $p_T > 10$\,GeV for events collected at $\sqrt{s} = 13$\,TeV, and $p_T > 15$\,GeV for events collected at $\sqrt{s} = 13.6$\,TeV. Jets are reconstructed using the anti-$k_T$ algorithm with $R=0.4$, requiring $p_T^j>25$\,GeV and $|\eta_j|<2.5$. Events with at least three reconstructed leptons are constructed and the $p_T$ of the leading lepton must have $p_T > 25$\,GeV and the subleading one $p_T > 15$\,GeV. An opposite-sign, same-flavor (OSSF) lepton pair is identified as a $Z$ if its invariant mass $|m_{\ell\ell} - m_Z| < 15$\,GeV. The signal region (SR) is divided into two parts, depending on the number of leptons: events with three leptons enter the ${\rm SR}_{3\ell}$, while events with four leptons are assigned to the ${\rm SR}_{4\ell}$. The ${\rm SR}_{3\ell}$ signal region also requires at least two jets with one of them to be $b$-tagged, whereas the ${\rm SR}_{4\ell}$ requires at least one $b$-tagged jet. 
\begin{equation}
\begin{tabular}{c|c|cc}
$tWZ$ SM & Run~2 + Run~3   \\
& [200 $\text{fb}^{-1}$]     \\
\hline
CMS &  134 \\
MPD &  70\\
\end{tabular}
\end{equation}

For Run~2 and Run~3 simulations we have considered $\sqrt{s} = 13$\,TeV and $13.6$\,TeV with integrated luminosities $138\,{\rm fb}^{-1}$ and $62\,{\rm fb}^{-1}$, respectively. The ratios for the number of events from the full detector simulations of CMS vs our fast simulation are 
\begin{equation}
\begin{aligned}
\frac{N_{\rm SM}^{\rm CMS}}{N_{\rm SM}^{\rm MPD}} &\approx  1.9\; ({\rm R2 + R3})
\end{aligned}
\end{equation}
which is reasonable because of the high multiplicity final state with of $3\ell$ and $4\ell$ and additional $(b)$-jets. Note that in the NP analysis, we will always consider the ratio of the NP signal strength over that of our SM simulation, such that detector effects drop out to a good approximation

\subsection{$ttZ$}

Here we reuse the results of our previous work, Ref~\cite{Ashanujjaman:2025una}, where we recasted $t\bar{t}Z$ analyses of ATLAS~\cite{ATLAS:2023eld} and CMS~\cite{CMS:2024mke} to search for signs of charged Higgs bosons and set novel limits on the product of branching fractions ${\rm Br}(t \to \Delta^\pm b)\times {\rm Br}(\Delta^\pm \to W Z)$. In Fig \ref{fig:Feynman_ttZ_tWZ}, we showed  the NP Feynman diagram contributing to the SM $t\bar{t}Z$ process. 

The CMS analysis provides differential cross sections for the sum of $t\bar{t}Z$ and $tWZ$ production (within the SM), unfolded to the parton level (after radiation but before hadronization), as functions of 5 observable. The ATLAS analysis, on the other hand, reports $t\bar{t}Z$ differential cross sections unfolded to both particle and parton
levels covering 15 observables. For the detailed information about the CMS and ATLAS observable, see \cite{CMS:2024mke, ATLAS:2023eld, Ashanujjaman:2025una}. The NP signal process $p p \to t \bar{t} \to W^\mp b \Delta^\pm b$ is simulated
for various  $m_{\Delta^\pm}$ in the 100–160\,GeV range. For the reconstruction and selection of physics objects, namely leptons (electrons and muons) and jets (including $b$-tagged jets), we closely follow the respective CMS and ATLAS analyses. We then performed a chi-square analysis with the available binned observed data, SM prediction, and the NP contribution, assuming ${\rm Br}(t \to \Delta^\pm b)\times {\rm Br}(\Delta^\pm \to W Z)$ as the NP signal strength and obtained the model-independent limit on ${\rm Br}(t \to \Delta^\pm b)\times {\rm Br}(\Delta^\pm \to W Z)$ (see left panel of Figure\,2 of Ref. \cite{Ashanujjaman:2025una}). 

%%%%%%%%%%%%%%%
\section{Model Description: The Real Higgs Triplet Model ($\Delta$SM)}

We consider the extension of the SM scalar sector by a real $\text{SU}(2)_L$ Higgs triplet with zero hypercharge ($Y=0$), commonly referred to as the $\Delta$SM~\cite{Ross:1975fq,Gunion:1989ci,Chankowski:2006hs,Blank:1997qa,Forshaw:2003kh,Chen:2006pb,Chivukula:2007koj,Bandyopadhyay:2020otm}. The model introduces a CP-even neutral scalar ($\Delta^0$) and a charged Higgs bosons ($\Delta^\pm$).

The SM Higgs doublet is represented by $\Phi$ and the real triplet by $\Delta$, can be decomposed as
\begin{align}
\Phi &= 
\begin{pmatrix}
\phi^+ \\
\frac{1}{\sqrt{2}}(v_\phi + h^0_\phi + i G^0)
\end{pmatrix}, \\
\quad
\Delta &= \frac{1}{2} 
\begin{pmatrix}
v_\Delta + h^0_\Delta & \sqrt{2} \Delta^+ \\
\sqrt{2} \Delta^- & - (v_\Delta + h^0_\Delta)
\end{pmatrix}.
\end{align}
The tree-level scalar potential is given by
\begin{align}
V &= -\mu^2_\phi\, \Phi^\dagger \Phi + \frac{\lambda_\phi}{4} (\Phi^\dagger \Phi)^2 - \mu^2_\Delta\, \mathrm{Tr}(\Delta^\dagger \Delta) \nonumber \\
&\quad + \frac{\lambda_\Delta}{4} [\mathrm{Tr}(\Delta^\dagger \Delta)]^2  + A\, \Phi^\dagger \Delta \Phi + \lambda_{\phi\Delta}\, \Phi^\dagger \Phi \cdot \mathrm{Tr}(\Delta^\dagger \Delta),
\end{align}
where all parameters are taken to be real, such that the potential is $CP$ conserving. {Note that in the limit $A \to 0$, the potential possesses a global
$O(4)_\Phi \times O(3)_\Delta$ symmetry and the discrete $Z_{2,\Delta} (\Delta \to -\Delta)$ symmetry. Therefore, a non-zero
$A$, corresponding to $A\Phi^\dagger\Delta\Phi$ term, leads to a soft breaking of this symmetry such that small values of it are natural in the t' Hooft sense and governs the mixing between the doublet and triplet scalars \cite{Ashanujjaman:2025una}}.

After electroweak symmetry breaking, the physical scalar spectrum consists of the SM-like Higgs boson $h$, identified with the 125\,GeV scalar, the neutral triplet-like scalar $\Delta^0$, and the charged scalars $\Delta^\pm$. The masses of the triplet-like scalars are approximately
\begin{align}
m^2_{\Delta^0} &\approx \frac{1}{2} \lambda_\Delta v_\Delta^2 + \frac{A v_\phi^2}{4 v_\Delta}, \\
m^2_{\Delta^\pm} &= \frac{A (v_\phi^2 + 4 v_\Delta^2)}{4 v_\Delta},
\end{align}
in an expansion in $v_\Delta/v$. The mixing angle $\alpha$ between the neutral scalars, defined as $\tan\alpha \approx \frac{4v_\Delta}{v}$, governs the deviation of Higgs couplings from their SM expectations and is constrained by precision Higgs measurements. 

The triplet vacuum expectation value $v_\Delta$ contributes at tree level only to the $W$ boson mass but not to the $Z$ mass, leading to a deviation in the $\rho$-parameter given by
\begin{align}
\rho_{\mathrm{tree}} \approx 1 + \frac{4 v_\Delta^2}{v^2},
\end{align}
where $v^2 = v_\phi^2 + 4 v_\Delta^2 \approx (246\,\mathrm{GeV})^2$. Electroweak precision data tightly constrain this contribution, requiring $v_\Delta <  \mathcal{O}(1\,\mathrm{GeV})$. Note that the model predicts a positive shift in the $W$-boson mass, potentially explaining the recently reported experimental deviation from the SM expectation. {Again the small mixing resulting from the $\mathcal{O}(1\,\mathrm{GeV})$ triplet vev, makes the  doublet
and triplet mostly decoupled and hence therefore free of tree-level flavor-changing effects \cite{Ashanujjaman:2024lnr}.}

To ensure theoretical consistency, the parameter space is subject to several constraints. Vacuum stability requires the scalar potential to be bounded from below in all field directions, which imposes the conditions
\begin{align}
\lambda_\phi > 0, \quad \lambda_\Delta > 0, \quad \sqrt{2}\, \lambda_{\phi\Delta} + \sqrt{\lambda_\phi \lambda_\Delta} > 0.
\end{align}
Perturbative unitarity bounds on all $2 \to 2$ scalar scattering amplitudes further constrain the quartic couplings, for instance,
\begin{align}
|\lambda_\phi| \leq 2 \kappa \pi, \quad |\lambda_\Delta| \leq 2 \kappa \pi, \quad |\lambda_{\phi\Delta}| \leq \kappa \pi, \quad \text{with } \kappa = 16.
\end{align}
These conditions restrict the allowed mass splitting between $\Delta^0$ and $\Delta^\pm$ to be only a few GeV once the bound on $v_\Delta$ from the $W$ mass is taking into account.

Note that since the triplet does not couple to the SM fermions at the renormalizable level, so the Yukawa Lagrangian remains SM-like and deviations of the SM Higgs couplings to fermion originate from mixing only.

Since $v_\Delta$ and thus the mixing with the SM Higgs is small, 
the triplet-like scalars are primarily produced at the LHC via Drell-Yan processes~\cite{FileviezPerez:2008bj},
\begin{align}
q \bar{q} \to Z^*/\gamma^* \to \Delta^\pm\Delta^\mp, \qquad q \bar{q}' \to W^{\pm *} \to \Delta^0 \Delta^\pm.
\end{align}

The neutral scalar $\Delta^0$ predominantly decays into $WW$, with loop-induced channels such as $\gamma \gamma$ and $Z \gamma$ becoming relevant when the scalar is nearly degenerate with $\Delta^\pm$. In Fig~\ref{fig:Br_150_neutral}, we show the branching ratios of $\Delta^0$ to various possible modes with respect to the scalar mixing angle $\alpha$. For $\alpha \simeq 0$, the ${\rm Br}(\Delta^0 \to WW)$ is $100\%$, whereas when ${\rm Br}(\Delta^0 \to WW) \approx 0$ (which corresponds to $\alpha \simeq 0.065$), other decay mode dominates. From the Fig~\ref{fig:Br_150_neutral}, at $m_{\Delta^\pm} = 150$ GeV and ${\rm Br}(\Delta^0 \to WW) \approx 0$, the dominant decays modes are $bb, ZZ$ with ${\rm Br}(\Delta^0 \to bb) \approx 50\%$ and ${\rm Br}(\Delta^0 \to ZZ) \approx 30\%$ respectively. 

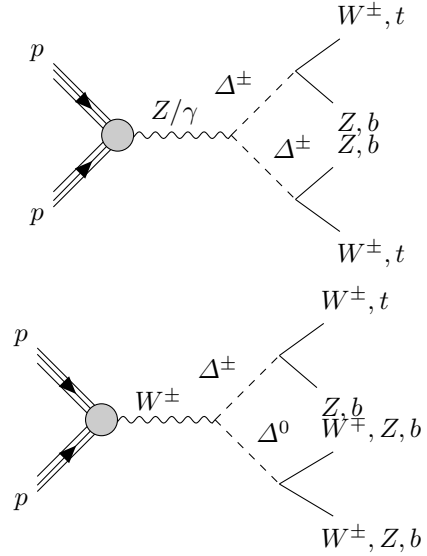
\begin{figure}[t!]
\centering
\begin{tikzpicture}[baseline=(current bounding box.center)]
\begin{feynman}
\vertex (a);
\vertex [above=0.1cm of a] (au);
\vertex [below=0.1cm of a] (ad);
\vertex [above left=1.2cm of a] (p1) {$p$}; 
\vertex [above left=1.2cm of au] (p1u); 
\vertex [above left=1.2cm of ad] (p1d);

\vertex [below left=1.2cm of a] (p2) {$p$}; 
\vertex [below left=1.2cm of au] (p2u); 
\vertex [below left=1.2cm of ad] (p2d);
\vertex [right=1.5cm of a] (b) ;

\vertex [above right=1.2cm of b] (e);
\vertex [below right=1.2cm of b] (f);

\vertex [above right=0.6cm of e] (ju) {$W^\pm,t$};
\vertex [below right=0.6cm of e] (jd) {$Z,b$};
\vertex [above right=0.6cm of f] (ku) {$Z,b$};
\vertex [below right=0.6cm of f] (kd){$W^\pm,t$};

\diagram{
(p1) -- [fermion] (a), (p1u) -- (au), (p1d) -- (ad), (p2) -- [fermion] (a) , (p2u) -- (au), (p2d) -- (ad), (a) -- [boson, edge label=${Z/\gamma}$] (b), (b) -- [scalar, edge label=$\Delta^\pm$] (e), (b) -- [scalar, edge label=$\Delta^\pm$] (f),
(e) --  (ju), (e) --  (jd), (f) --  (ku), (f) --  (kd)
};
\end{feynman}
\node[draw, circle, minimum size=12pt, inner sep=0pt, fill=gray!40] at (a) {};
\end{tikzpicture}

\begin{tikzpicture}[baseline=(current bounding box.center)]
\begin{feynman}
\vertex (a);
\vertex [above=0.1cm of a] (au);
\vertex [below=0.1cm of a] (ad);
\vertex [above left=1.2cm of a] (p1) {$p$}; 
\vertex [above left=1.2cm of au] (p1u); 
\vertex [above left=1.2cm of ad] (p1d);

\vertex [below left=1.2cm of a] (p2) {$p$}; 
\vertex [below left=1.2cm of au] (p2u); 
\vertex [below left=1.2cm of ad] (p2d);
\vertex [right=1.5cm of a] (b) ;

\vertex [above right=1.2cm of b] (e);
\vertex [below right=1.2cm of b] (f);

\vertex [above right=0.6cm of e] (ju) {$W^\pm,t$};
\vertex [below right=0.6cm of e] (jd) {$Z,b$};
\vertex [above right=0.6cm of f] (ku) {$W^\mp,Z,b$};
\vertex [below right=0.6cm of f] (kd){$W^\pm,Z,b$};

\diagram{
(p1) -- [fermion] (a), (p1u) -- (au), (p1d) -- (ad), (p2) -- [fermion] (a) , (p2u) -- (au), (p2d) -- (ad), (a) -- [boson, edge label=${W^\pm}$] (b), (b) -- [scalar, edge label=$\Delta^\pm$] (e), (b) -- [scalar, edge label=$\Delta^0$] (f),
(e) --  (ju), (e) -- (jd), (f) --  (ku), (f) -- (kd)
};
\end{feynman}
\node[draw, circle, minimum size=12pt, inner sep=0pt, fill=gray!40] at (a) {};
\end{tikzpicture}
\caption{Feynman diagram for the Drell-Yan production of triplet scalars $pp \to \Delta^\pm\Delta^\mp, \Delta^\pm \Delta^0$ further decaying to $W,Z$, which will contribute to SM $VVV$ processes, where $V = W,Z$.}
\label{fig:Feynman_VVV}
\end{figure}

The charged scalar $\Delta^\pm$ typically decays to $WZ$, $\tau \nu$, or $tb$, depending on its masses. In Fig.~\ref{fig:Br_150_neutral}, we have shown the branching ratios of $\Delta^\pm$ as a function of its mass. 

In Fig.~\ref{fig:Br_150_neutral}, the uncertainties are estimated by propagating the errors on the $\tau\tau , c\bar{c}, t\bar{t}, WW$ and $ZZ$ decays of a hypothetical SM-like Higgs with mass $m_{\Delta^0}$ and $m_{\Delta^\pm}$, respectively, as reported in the CERN Yellow Report \cite{LHCHiggsCrossSectionWorkingGroup:2013rie}.

The phenomenological implications in di-photon final states were examined in Refs.~\cite{Ashanujjaman:2024pky,Crivellin:2024uhc,Ashanujjaman:2024lnr}, finding that the $\Delta$SM can account for the excess at 152\,GeV. We also have recasted $t\bar{t}Z$ analyses of LHC~\cite{ATLAS:2023eld,CMS:2024mke} to search for signs of charged Higgs bosons  and set novel limits on the product of branching fractions ${\rm Br}(t \to \Delta^\pm b)\times {\rm Br}(\Delta^\pm \to W Z)$, which when translated onto the $\Delta$SM shows a $\sim 2\sigma$ preference, see Ref~\cite{Ashanujjaman:2025una}. Multi-lepton searches we studied in Refs.~\cite{Crivellin:2021ubm,Ashanujjaman:2024lnr}, finding that the scalar triplet is consistent with data. Here we want to consider these channels in more detail recasting the multi-boson searches discussed above.

\section{Results}

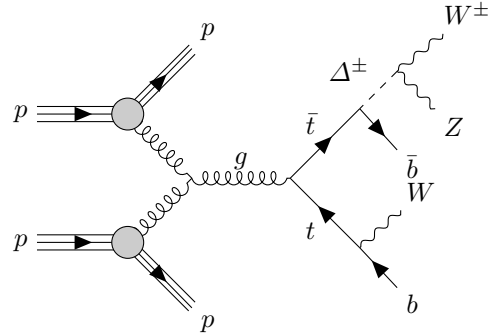
\begin{figure}[t!]
\centering
\begin{tikzpicture}[baseline=(current bounding box.center)]
\begin{feynman}
% interaction point
\vertex (a);
% gluons from protons
\vertex [above left=1.2cm of a] (c);
\vertex [above=0.1cm of c] (cu); 
\vertex [below=0.1cm of c] (cd);

\vertex [below left=1.2cm of a] (d); 
\vertex [above=0.1cm of d] (du); 
\vertex [below=0.1cm of d] (dd);
% incoming protons
\vertex [left=1.2cm of c] (p1) {$p$}; 
\vertex [left=1.2cm of cu] (p1u); 
\vertex [left=1.2cm of cd] (p1d);

\vertex [left=1.2cm of d] (p2) {$p$}; 
\vertex [left=1.2cm of du] (p2u); 
\vertex [left=1.2cm of dd] (p2d);

% outgoing protons
\vertex [above right=1.2cm of c] (pp1) {$p$}; 
\vertex [above right=1.15cm of cu] (pp1u); 
\vertex [above right=1.26cm of cd] (pp1d); 

\vertex [below right=1.2cm of d] (pp2) {$p$}; 
\vertex [below right=1.26cm of du] (pp2u); 
\vertex [below right=1.14cm of dd] (pp2d); 

% central propagator
\vertex [right=1.3cm of a] (b) ;

% outgoing top pair
\vertex [above right=1.3cm of b] (e);
\vertex [below right=1.3cm of b] (f);

% H+ branch
\vertex [above right=0.7cm of e] (j); 
\vertex [above right=0.7cm of j] (m){$W^\pm$};
\vertex [below right=0.7cm of j] (n){$Z$};
\vertex [below right=0.7cm of e] (i) {$\bar b$};

% other top decay
\vertex [above right=0.7cm of f] (k) {$W$};
\vertex [below right=0.7cm of f] (l) {$b$};

\diagram{
% protons to gluons
(p1) -- [fermion] (c) -- [gluon] (a), (p1u) -- (cu), (p1d) --  (cd),
(p2) -- [fermion] (d) -- [gluon] (a),  (p2u) -- (du), (p2d) -- (dd),

% gluons to protons
(c) -- [fermion] (pp1), (cu) -- (pp1u), (cd) -- (pp1d),
(d) -- [fermion] (pp2), (du) -- (pp2u), (dd) -- (pp2d),

% gluon fusion
(a) -- [gluon, edge label=$g$] (b),

% top pair
(f) -- [fermion, edge label=$t$] 
(b) -- [fermion, edge label=$\bar t$] (e),

% top -> \Delta+ b
(j) -- [scalar, edge label'=$\Delta^\pm$] (e) -- [fermion] (i),
(l) -- [fermion] (f) -- [boson] (k),

% H+ decay
(j) -- [boson] (m),
(j) -- [boson] (n)
};
\end{feynman}
\node[draw, circle, minimum size=12pt, inner sep=0pt, fill=gray!40] at (c) {};
\node[draw, circle, minimum size=12pt, inner sep=0pt, fill=gray!40] at (d) {};
\end{tikzpicture}
\caption{Representative Feynman diagram for $pp \to t\bar{t}$ with $t \to \Delta^\pm b$ and $\Delta^\pm \to W^\pm Z$, leading to a $t\bar{t}Z-$like SM signature and $tWZ-$like SM signature with additional $b$-jets.}
\label{fig:Feynman_ttZ_tWZ}
\end{figure}

\begin{table*}[h!]
	\centering
	\begin{tabular}{l | c c c c c c }
		\hline
		\textbf{Analysis} &  $\mu_{\rm EXP} = \sigma_{\rm EXP}/\sigma_{\rm SM}$ &$\mu_{\Delta^\pm\Delta^0}$ &  $\mu_{\Delta^\pm\Delta^\mp}$& $\mu_{t\bar{t}}$ & $\mu_{\rm NP}$ \\
        & & {\small  $WW=100\%(WW=0\%)$}  & & &\\
		\hline
		$VVZ$~\cite{ATLAS:2024nab} & 1.43 $\pm$ 0.29  & 0.33(0.15) & 0.19 & 0 & 0.52(0.34)\\
        $WWW$~\cite{ATLAS:2022xnu} & 1.61 $\pm$ 0.25  & 0.76(0.07) & 0.14 & 0 & 0.90(0.21)\\
        $WWZ$(Run~2)~\cite{CMS:2025hlu} & 0.52 $\pm$ 0.52 & 0.86(0.63) & 0.64 & 0& 1.50(1.27)\\
        $WWZ$(Run~3)~\cite{CMS:2025hlu} & 1.91 $\pm$ 1.15 & 0.99(0.68) & 0.53 & 0 & 1.52(1.21)\\
        $tWZ$(Run~2)~\cite{CMS:2025xph} & 1.82 $\pm$ 0.38  & 0.20(0.41) & 0.25 & 0.93 & 0.93(1.05)\\
        $tWZ$(Run~3)~\cite{CMS:2025xph} & 1.64 $\pm$ 0.52 & 0.19(0.38) &  0.22 & 0.73 & 0.79(0.90) \\
        \hline
        % $ttZ$(CMS)~\cite{CMS:2024mke} & 1.24 $\pm$ 0.32  & 0  & 0 & 0.20 & 0.20\\
        % $ttZ$(ATLAS)~\cite{ATLAS:2023eld} & 1.05 $\pm$ 0.30 &  0 & 0 & 0.08 & 0.08 \\
		\hline
	\end{tabular}
    \caption{Predicted NP signal strength relative to the SM ($\mu_{\rm NP}$), along with the contribution from the relevant NP processes: $pp\to W^+\to \Delta^\pm\Delta^0$($\mu_{\Delta^\pm\Delta^0}$), $pp\to \gamma^*,Z^*\to \Delta^\pm \Delta^\mp$ ($\mu_{\Delta^\pm\Delta^\mp}$) and $pp \to t\bar{t}\to\Delta^\pm b Wb$ ($\mu_{t\bar{t}}$). The NP contribution from $t\bar{t}$ channel for $tWZ$ is normalized to Br$(t\to \Delta^\pm b)=10^{-3}$. }
    \label{tab:Np_signalstregth}
\end{table*}

In the $\Delta$SM we have two different processes which lead to the production of EW gauge bosons and thus multiple leptons. First, there is the Drell-Yan production of $\Delta^\pm \Delta^0$ and $\Delta^\pm \Delta^\mp$ shown in Fig.~\ref{fig:Feynman_VVV}. This effect is unavoidable and will be discussed in the next subsection. In addition, we have the process $pp\to t\bar t\to \Delta^\pm b W \bar b$ which can happen in our setup with $m_{\Delta^\pm}\approx 150$\,GeV. However, this rate depends on $v_\Delta$ and thus involves an additional free parameter.

For the simulation, we followed the same chain of packages as for the SM. Furthermore, to correct for the effects of our fast simulation, we will always consider the ratio of NP over SM, i.e.~$\mu_{\rm NP}=\sigma^{\rm vis}_{\Delta{\rm SM}}/\sigma^{\rm vis}_{\rm SM}$, where vis stands for the visible cross section (i.e.~after acceptance, efficiency and cuts).

Therefore, this $\chi^2$ function can be constructed as
\begin{equation}
    % \chi^2 = \Bigg(\frac{\mu_{\rm EXP} - \Big(1 + x \times (\mu_{\Delta^\pm\Delta^\mp} + \mu_{\Delta^\pm\Delta^0}) +{\rm Br}\times \mu_{t\bar{t}} \Big)}{\Delta \mu}\Bigg)^2
    \chi^2 = \Bigg(\frac{\mu_{\rm EXP} - \Big(1 + x \times \mu_{\rm NP} \Big)}{\Delta \mu}\Bigg)^2.
    \label{eq:chisqr}
\end{equation}
Here, we introduced the fitting parameter $x$ {($x$ is not a physical parameter of the model, rather an effective scaling factor corresponding to the new physics signal strength)} such that the NP signal strength can be varied for illustrational purposes. The $\Delta$SM scenario corresponds to $x =1$. 

\subsection{$pp\to \Delta^\pm \Delta^0$ and $pp\to \Delta^\pm \Delta^\mp$}

Fig.~\ref{fig:Feynman_VVV}, shows the DY production of the triplet scalars $pp \to \Delta^\pm \Delta^0, \Delta^\pm\Delta^\mp$ with their main decay modes into gauge bosons. While the branching ratios of $\Delta^\pm$ are determined as a function of the mass only, the dominant decay modes of $\Delta^0$ depend in addition on the mixing angle $\alpha$ with the SM Higgs (see Fig.~\ref{fig:Br_150_neutral}). Therefore, we will consider two limiting case for the latter: 1) Br$(\Delta^\pm\to WW)\approx 100\%$ and 2) Br$(\Delta^\pm\to WW)\approx 0\%$, which leads to $(\Delta^\pm\to ZZ)\approx 30\%$ and $(\Delta^\pm\to bb)\approx 50\%$.

\textbf{\textit{VVZ:}} The NP contribution to the SM $pp \to VVZ$ process, where $V = W,Z$, comes from the $\Delta^\pm\Delta^\mp$ channel and the $\Delta^\pm\Delta^0$ channel, the latter depending on the Br$(\Delta^0 \to WW)$. The visible NP cross section originating from the $pp \to \Delta^\mp\Delta^\pm$ is 0.57\,fb. The process $pp \to \Delta^0\Delta^\pm$ contribution is 0.99 (0.45)\,fb for  ${\rm Br}(\Delta^0\to WW)=$100\% (0\%). This results in a NP signal strength relative to the SM ($\mu_{\rm NP}^{VVZ}$) of 0.52(0.34) for  ${\rm Br}(\Delta^0\to WW)=100\% (0\%)$.

\textbf{\textit{WWW:}} The DY pair production NP contribution to the $pp \to WWW$ process is 0.27\,fb from $pp \to \Delta^\mp\Delta^\pm$ and 1.47(0.14)\,fb from $pp \to \Delta^0\Delta^\pm$ with ${\rm Br}(\Delta^0\to WW)=100\% (0\%)$. The NP signal strength relative to the SM ($\mu_{\rm NP}^{WWW}$) for ${\rm Br}(\Delta^0\to WW)=100\%(0\%)$ is 0.90(0.21).

\textbf{\textit{WWZ:}} The NP contribution to the SM $pp \to WWZ$ process for Run~2 is 0.04\,fb from $pp \to \Delta^\mp\Delta^\pm$ and 0.053 (0.039)\,fb from $pp \to \Delta^0\Delta^\pm$ with ${\rm Br}(\Delta^0\to WW)=100\% (0\%)$ and for Run~3 the contributions are 0.03\,fb and 0.057(0.040)\,fb, respectively. The NP signal strength relative to the SM ($\mu_{\rm NP}^{WWZ}$) for ${\rm Br}(\Delta^0\to WW)=100\%(0\%)$ for Run~2 is 1.50(1.27) and for Run~3 is 1.52(1.21).

\begin{figure*}[htb!]
\centering
\begin{subfigure}{0.45\textwidth}
    \includegraphics[width=0.9\linewidth, height=6.5cm]{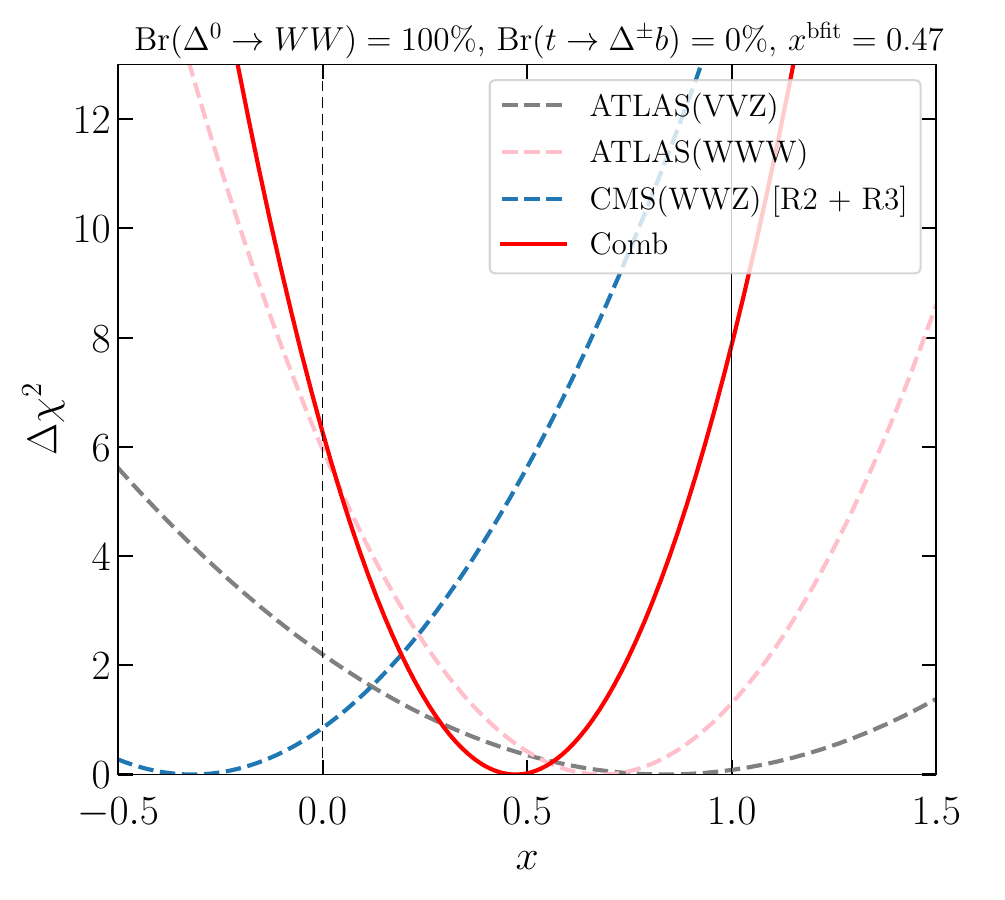} 
\end{subfigure}
\begin{subfigure}{0.45\textwidth}
    \includegraphics[width=0.9\linewidth, height=6.5cm]{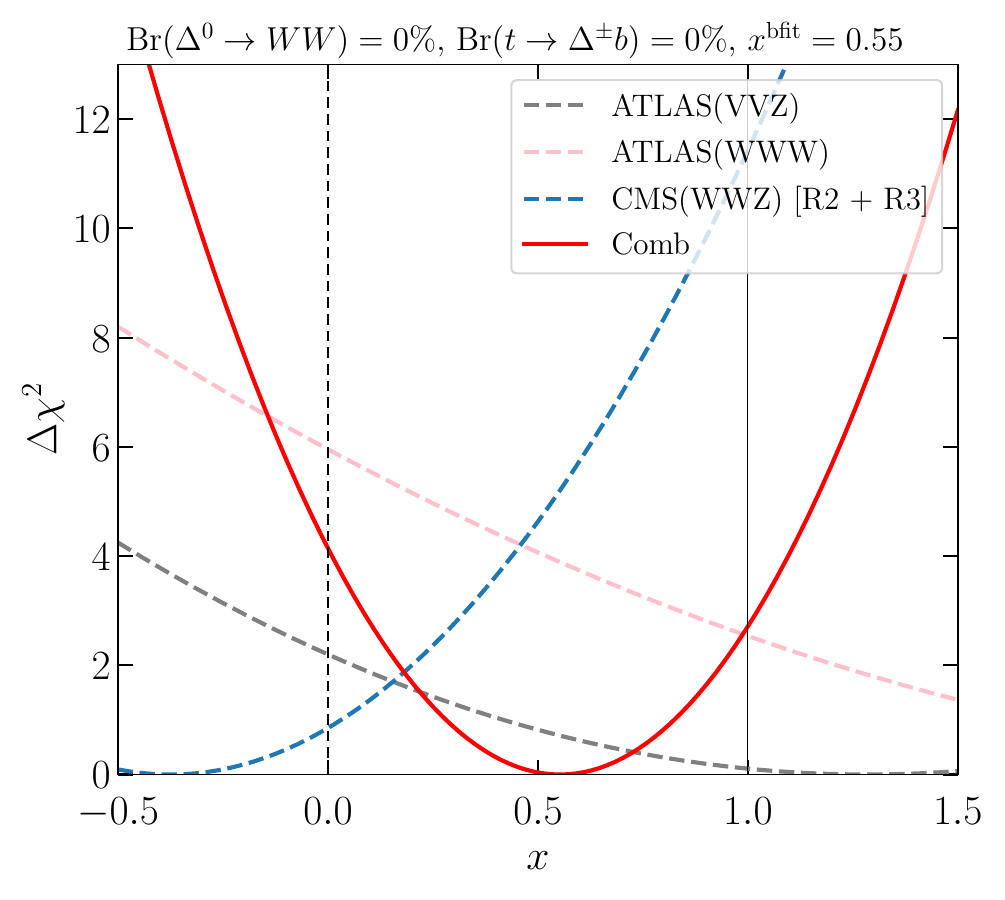}
\end{subfigure}
\caption{Left: Chi-square distribution as a function of the fitting parameter $x$ for the tri-boson channels (excluding $tWZ$). We consider for ${\rm Br}(\Delta^0 \to WW) = 100\%$ and ${\rm Br}(t \to \Delta^\pm b) = 0\%$. The red line represents the combined chi square distribution, with $\chi^2_{\rm comb}(x = 0) =  15.8$, $\chi^2_{\rm comb}({\rm min}) =  9.5$, and deviation w.r.t.~the SM: $\sqrt{\Delta\chi^2_{\rm comb}(x = 0)} \approx  2.5\sigma$. Right: Same for ${\rm Br}(\Delta^0 \to WW) = 0\%$ resulting in $\chi^2_{\rm comb}({\rm min}) = 11.6$ corresponding to $\sqrt{\Delta\chi^2_{\rm comb}(x = 0)} \approx 2\sigma$.}
\label{fig:Chisqr_Br0}
\end{figure*}

\textbf{\textit{tWZ:}} The DY processes $pp \to \Delta^\mp\Delta^\pm$ and $pp \to \Delta^\mp\Delta^0$ can also contribute to SM process $pp \to tWZ$, which results in a $WWZb$ final state. From $\Delta^\mp\Delta^\pm$ process, when one of the $\Delta^\pm$ decays to $t^*b(Wbb)$ and others decays to $WZ$, we can have $WWZb$ with additional $b$-jets, where the ${\rm Br}(\Delta^\pm \to WZ) \approx 75\%$, and ${\rm Br}(\Delta^\pm \to tb) \approx 17\%$ (see Fig. \ref{fig:Br_150_neutral}). Similarly, from $\Delta^\mp\Delta^0$ process, we can also get $tWZ$ like signature with additional jets when $\Delta^\pm \to WZ,t^*b$ and $\Delta^0 \to WW,ZZ,bb$. The $\Delta^\mp\Delta^0$ contribution is considered in two extreme situations depending upon the ${\rm Br}(\Delta^0 \to WW)$: $WW=100\%$ and  $WW=0\%(ZZ \approx 30\%{\rm ,and~} bb \approx 50\%)$. The Run~2 $\Delta^\mp\Delta^\pm$ contribution to SM $tWZ$ process is 0.09\,fb and $\Delta^\pm\Delta^0$ contribution for $WW = 100\%(WW = 0\%)$ is 0.071(0.14)\,fb. Similarly for Run~3 the contributions are 0.076\,fb and 0.069(0.13)\,fb, respectively.

\textbf{\textit{ttZ:}} Similar to the $tWZ$ process, the $pp \to \Delta^\pm\Delta^\mp$ could contribute to $ttZ$ like-final states. However, in this case, the SM cross section is significantly larger than the $tWZ$ cross section and the efficiency of the cuts for the NP signal $\Delta^\pm\Delta^\mp\to WZ t^*b$ is lower: We found that the relative NP signal strength compared to the SM is below 1\%, both for ATLAS and CMS. Therefore, we can neglect the Drell-Yan contribution to the $ttZ$ channel and will only include the effect of $pp\to t\bar{t}\to \Delta^\pm b W b$ for which we can directly reuse the results of Ref.~\cite{Ashanujjaman:2025una} (see next subsection).

\textbf{\textit{Drell-Yan combination:}}  The resulting $\chi^2$ as a function of the fitting parameter $x$ is given in Fig.~\ref{fig:Chisqr_Br0}, where we have assumed ${\rm Br}(t \to \Delta^\pm b)=0$ and disregarded the small effect in the $ttZ$ signal region. In left (right) panel of Fig.~\ref{fig:Chisqr_Br0}, we show the results for ${\rm Br}(\Delta^0 \to WW = 100\%)$ (${\rm Br}(\Delta^0 \to WW = 0\%)$). At $x = 0$, the combined chi square ($\chi^2_{\rm comb}$) is $\chi^2_{\rm comb}(x=0) = 15.74$, and the minimum of the $\chi^2_{\rm comb}$ for $100\% (0\%) $ $WW$ branching is $\chi^2_{\rm comb}(x_{\rm min}) = 9.47(11.6)$ for best fit value of the fitting parameter $x^{\rm b-fit} = 0.47 (0.55)$ respectively. Therefore, we have a $\Delta \chi^2=\chi^2_{\rm comb}(x = 0) - \chi^2_{\rm comb}({\rm min}) = 6.27(4.14)$ corresponding to $\sqrt{\Delta \chi^2_{\rm comb}(x = 0)} = 2.5\sigma(2.03\sigma)$, respectively. However, the $\Delta$SM predicts $x=1$, with a $\chi^2$ of 17.34(14.31), such that the model is consistent with data but not preferred over the SM hypothesis.

\subsection{$pp\to t\bar t\to \Delta^\pm b Wb$}

Fig.~\ref{fig:Feynman_ttZ_tWZ} shows the contribution from the $\Delta$SM, which contributes to the $tWZ$ and $ttZ$ analyses. As we can neglect the effect of the Drell-Yan production in the $ttZ$ channel, we can directly reuse the results of Ref.~\cite{Ashanujjaman:2025una}:
\begin{equation}
{\rm Br}(\Delta^\pm \to W^\pm Z)=(0.13 \pm 0.064)\%\qquad {\rm (CMS)}\,,
\end{equation}
and
\begin{equation}
{\rm Br}(\Delta^\pm \to W^\pm Z)=(0.054 \pm 0.037)\%\qquad {\rm (ATLAS)}\,.
\end{equation}

In the $tWZ$ signal region, we have the contribution from the Drell-Yan production discussed previously and the effect of $pp \to t\bar{t} \to \Delta^\pm bWb$ (see Fig.~\ref{fig:Feynman_ttZ_tWZ}). The obtained NP signal strengths for the  $t\bar{t}$ process $ttZ$ Run 2 (Run 3) is \begin{equation}
    \mu_{t\bar{t}}^{ttZ}=0.93(0.73)\times{\rm Br}(t \to \Delta^\pm b)/10^{-3}\,.
\end{equation}

\begin{figure*}[htb!]
\centering
\begin{subfigure}{0.45\textwidth}
    \includegraphics[width=0.9\linewidth, height=6.5cm]{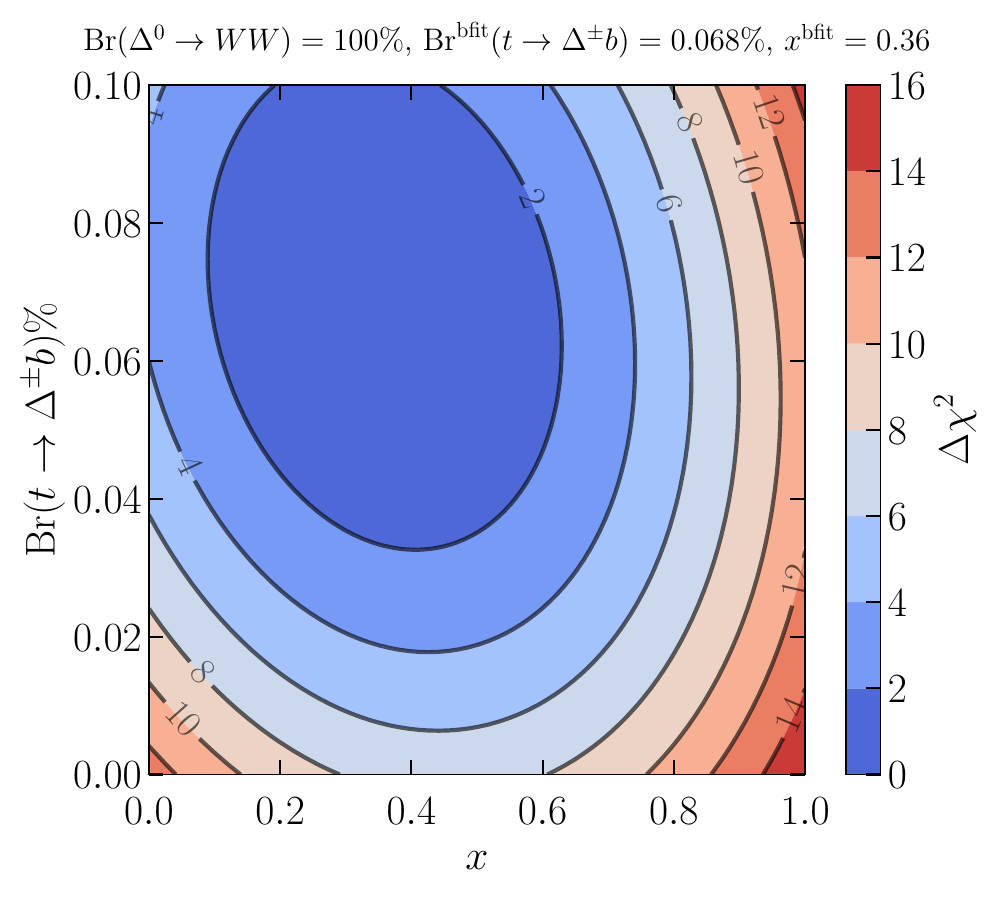} 
\end{subfigure}
\begin{subfigure}{0.45\textwidth}
    \includegraphics[width=0.9\linewidth, height=6.5cm]{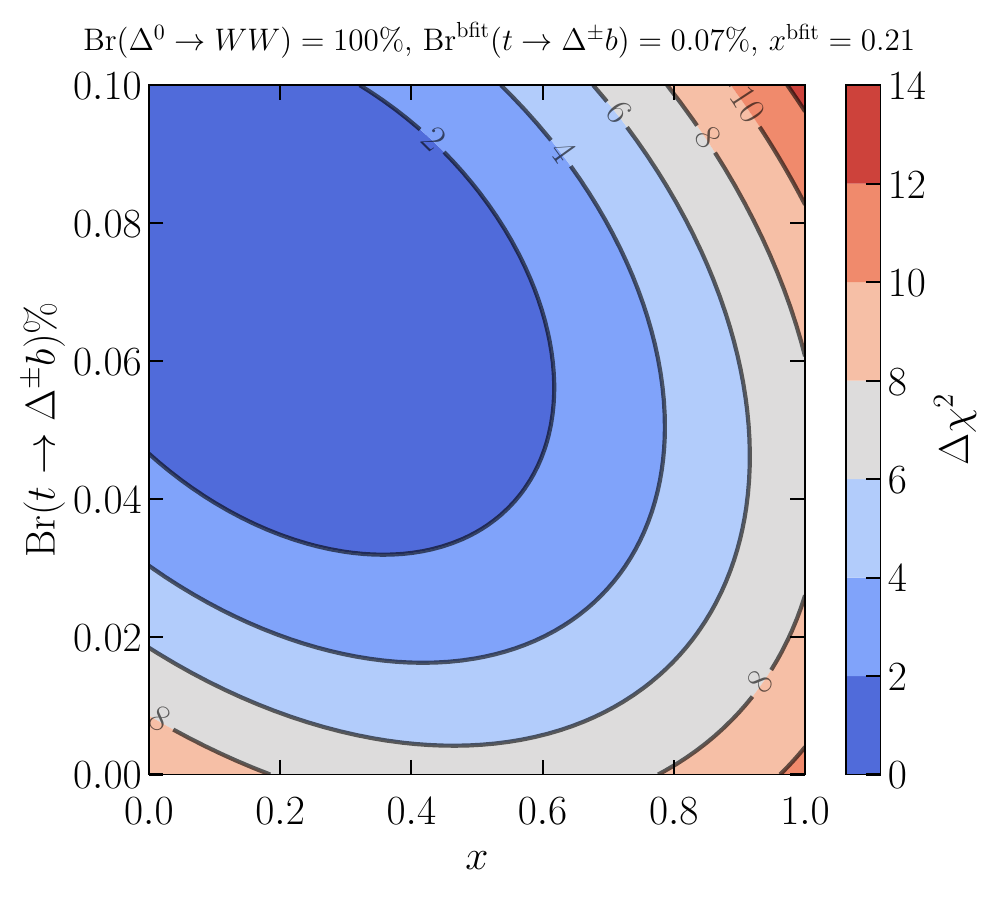}
\end{subfigure}
\caption{Left: $\chi^2$ contours as a function of $x$ and ${\rm Br}(t \to \Delta^\pm b)$ assuming ${\rm Br}(\Delta^0 \to WW) = 100\%$. We find a preference for NP of $ \approx 3 \sigma$ (for 2DoF). Same for ${\rm Br}(\Delta^0 \to WW) = 0\%$ resulting in  $\approx 2.5\sigma$. In the above chi-square distribution, $x = 0$ and ${\rm Br}(t \to \Delta^\pm b) = 0$ corresponds to the SM model, while $x=1$ corresponds to the real triplet model.}
\label{fig:Chisqr_dist_contour}
\end{figure*}

We can thus perform the combined fit to $x$ and  ${\rm Br}(t \to \Delta^\pm b)$ in Fig.~\ref{fig:Chisqr_dist_contour} for ${\rm Br}(\Delta^0 \to WW) = 100\%$ shown in the left and for ${\rm Br}(\Delta^0 \to WW) = 0\%$ shown in the right panel. From the left with $100\% WW$ mode, we get best fit values as ${\rm Br}^{\rm bfit}(t \to \Delta^\pm b) = 0.068\%$ and $x^{\rm bfit} = 0.36$. Similarly from the left with $0\% WW$ decay of $\Delta^0$, we get best fit values as ${\rm Br}^{\rm bfit}(t \to \Delta^\pm b) = 0.07\%$ and $x^{\rm bfit} = 0.21$. This corresponds to a preference over the SM hypothesis of $2.97\sigma$ and $2.47\sigma$ assuming 2 degree of freedom ($x$ and ${\rm Br}(t \to \Delta^\pm b)$), respectively. In this case, the $\Delta$SM with $x=1$ shows a slight preference over the SM of $1.3\sigma (1.6\sigma)$ and Br=0.053\% (0.043\%).

\section{Conclusion}
Motivated by the long-standing multi-lepton anomalies at the LHC, we previously argued for the presence of an additional scalar state with a mass around 150\,GeV. Subsequent LHC measurements in fact show indications for a narrow resonance with $152\pm1$\,GeV which can be partially explained with the $\Delta$SM model.

In this framework, the same dynamics that account for the observed $\gamma\gamma+X$ signatures imply correlated effects in triboson production. In particular, the model predicts an enhancement in $VVV$ rates data. While there are indications for enhanced tri-boson cross sections, the $\Delta$SM predicts a higher signal strength than the currently preferred central value, resulting in an inclusive result. Exploiting the available Run~2 measurements, we reported the first hint of the predicted excess in $VVV$ production, with a significance of $2.6\sigma$. With the increased luminosity and improved systematics expected from Run~3, the LHC program should be able to decisively test this prediction. Confirmation of the triboson excess would provide an independent and complementary validation of the emerging picture of new bosonic degrees of freedom in the LHC data, and would strongly motivate dedicated differential studies and global fits to pin down the underlying electroweak representation and couplings. 

Finally, we need to note that the charged component of the triplet can be scrutinised in great detail at future electron-positron colliders~\cite{Maharathy:2025yyo}. This includes precise measurements of the masses, and exploration of couplings and quantum numbers with precision that cannot be attained at high-energy hadron colliders. 

\appendix

% \section{Cutflow}
\section{Appendix}
This section contains the cut-flows performed for the analysis considered in the article, see Table.~[\ref{tab:xsections_sensitivity}, \ref{tab:Np_signalstregth}]. The cut flows for all the analysis is shown as, $VVZ$:  Table.~[\ref{tab:ATLASVVZ_3l},\ref{tab:ATLASVVZ_4l},\ref{tab:ATLASVVZ_5l}], $WWZ$(Run 2 and Run 3): Table.~[\ref{tab:CMSWWZ_R2}, \ref{tab:CMSWWZ_R3}], $WWW$: Table.~[\ref{tab:ATLASWWW_2l}, \ref{tab:ATLASWWW_3l}], and $tWW$(Run 2 and Run 3): Table.~[\ref{tab:CMStWZ_R2}, \ref{tab:CMStWZ_R3}].

\begin{table*}[h!]
	\centering
	\begin{tabular}{l c c c c}
		\hline
		\textbf{Cut} & $\Delta^0 \Delta^\pm$ & $\Delta^0 \Delta^\pm$ & $\Delta^\mp \Delta^\pm$  & SM $VVZ$ \\
         			&  $WW$(100\%)  & $WW$(0\%)  &  & \\
		\hline
		$\geq 3$ leptons 											& 537 & 185  & 287  & 655   \\
		Exactly 3 leptons and 1 jet 								& 430  & 145 & 238  & 539   \\
		Lepton $p_T$ cuts (27,15,15) 								& 327 & 112  & 182  & 476   \\
		$\geq 1$ OSSF pair 											& 290 &110  & 177  & 469   \\
		All OSSF $m_{\ell \ell} \geq 12 \text{GeV}$					& 277 &106   & 170  & 466   \\
		$Z$ mass window $|m_{\ell\ell} - M_Z| < 20$ GeV 				& 132 &69  & 88  & 445   \\
		$1.37 < |\eta_\ell| < 1.52$ 								& 132 &69  & 87  & 443   \\
		$\ge 1 \text{ jet with } p_T > 20~\text{GeV}$, $b$-jet veto   & 111 &48     & 63  & 366   \\
		$3\ell$-1j 													& 26 &12  & 14  & 77   \\
		$3\ell$-2j (in V) 											&31  &13  & 21  & 123    \\
		$3\ell$-2j (out V) 											& 54  &23 & 28  & 165   \\
		\hline
	\end{tabular}
    \caption{Cutflow table $VVZ$(ATLAS)~\cite{ATLAS:2024nab} $3\ell$ SR. }
    \label{tab:ATLASVVZ_3l}
\end{table*}

\begin{table*}[h!]
	\centering
	\begin{tabular}{l c c c c}
		\hline
		\textbf{Cut} & $\Delta^0 \Delta^\pm$ &$\Delta^0 \Delta^\pm$  & $\Delta^\mp
		\Delta^\pm$  & SM $VVZ$ \\
        &  $WW$(100\%) &  $WW$(0\%) &  & \\
		\hline
		$= 4$ leptons & 57 & 26 &  34 &71\\
		Lepton $p_T$ cuts (30,15,8,6) & 57 & 25 & 33 & 68\\
		$\geq 1$ OSSF pair & 56 & 25 & 33 & 68\\
		All OSSF $m_{\ell \ell} \geq 12 \text{GeV}$& 52 & 23 & 30 & 66\\
		Z mass window $|m_{\ell\ell} - M_Z| < 20$ GeV & 31 & 19 &  19 & 66\\
		$p_T^{miss} > 10 $\,GeV, $b$-jet veto  & 26 & 13 & 15 &55\\
		$4\ell$-DF & 12 & 2 &6 & 20\\
		$4\ell$-SF (in V) & 3 & 2 & 2 & 18\\
		$4\ell$-SF (out V) & 12 & 9 & 7 & 17\\
		\hline
	\end{tabular}
    \caption{Cutflow table $VVZ$(ATLAS)~\cite{ATLAS:2024nab} $4\ell$ SR.}
        \label{tab:ATLASVVZ_4l}
\end{table*}

\begin{table*}[h!]
	\centering
	\begin{tabular}{l c c c c}
		\hline
		\textbf{Cut} & $\Delta^0 \Delta^\pm$ & $\Delta^0 \Delta^\pm$  & $\Delta^\mp
		\Delta^\pm$ & SM $VVZ$ \\
        &  $WW$(100\%) & $WW$(0\%) &  & \\
		\hline
		$\geq 5$ leptons & 4.89  & 4.64 & 3.43 & 2.97\\
		$\geq 2$ OSSF pair & 1.49  & 1.96 & 1.48 & 2.97\\
		$b$-jet veto  & 1.49  & 1.76 & 1.43 & 2.71\\
		\hline
	\end{tabular}
    \caption{Cutflow table $VVZ$(ATLAS)~\cite{ATLAS:2024nab} $5\ell$ SR. %\red{more digits}
    }
    \label{tab:ATLASVVZ_5l}
\end{table*}

\begin{table*}[h!]
	\centering
	\begin{tabular}{l c c c c}
		\hline
		\textbf{Cut} & $\Delta^0 \Delta^\pm$ & $\Delta^0 \Delta^\pm$ &  $\Delta^\mp
		\Delta^\pm$ & SM $WWZ$ \\
        &  $WW$(100\%) &  $WW$(0\%)  &  & \\
		\hline
		$= 4$ leptons & 65.88 & 28.65 & 35.6 & 24.38 \\
        $p_T(\ell_1) > (25,15,10,10)$ & 65.36 & 28.45 & 35.4 & 24.38 \\
        $b$-jet veto & 13.51 & 9.36 & 8.85 & 13.36 \\
        OF & 5.55 & 2.93 & 3.23 & 5.74 \\
        SF & 1.78 & 2.46 & 2.24 & 2.64\\
		\hline
	\end{tabular}
    \caption{Cutflow table $WWZ$(CMS)~\cite{CMS:2025hlu} $4\ell$ SR for Run~2. %\red{These are too many digits. Keep in mind, we have an order 10\% accuracy. So two signigicant digits}
    }
    \label{tab:CMSWWZ_R2}
\end{table*}

\begin{table*}[h!]
	\centering
	\begin{tabular}{l c c c c}
		\hline
		\textbf{Cut} & $\Delta^0 \Delta^\pm$ & $\Delta^0 \Delta^\pm$  &  $\Delta^\mp
		\Delta^\pm$ & SM $WWZ$ \\
        &  $WW$(100\%) &  $WW$(0\%)  &  & \\
		\hline
		$= 4$ leptons & 27.01 & 12.74 & 15.41 & 10.84 \\
        $p_T(\ell_1) > (25,15,10,10)$ & 26.92 & 12.73 & 15.32 & 10.84 \\
        $b$-jet veto & 5.32 & 3.93 & 3.58 & 6.25 \\
        OF & 2.4 & 1.34 & 1.38 & 2.48 \\
        SF & 1.13 & 1.11 & 1.1 & 1.12\\
		\hline
	\end{tabular}
    \caption{Cutflow table $WWZ$(CMS)~\cite{CMS:2025hlu} $4\ell$ SR for Run~3.}
    \label{tab:CMSWWZ_R3}
\end{table*}

\begin{table*}[h!]
	\centering
	\begin{tabular}{l c c c c}
		\hline
		\textbf{Cut} & $\Delta^0 \Delta^\pm$ & $\Delta^0 \Delta^\pm$ & $\Delta^\mp
		\Delta^\pm$ & SM $WWW$ \\
        &  $WW$(100\%) &  $WW$(0\%) &  & \\
		\hline
		$= 2$ leptons & 2296 & 564 & 863 & 2931\\
		Leptons with Same charge & 643 & 87 & 143 & 838\\
		$\ge 1 \text{ jet }$, $b$-jet veto  & 362 & 46 & 81 & 498\\
		$p_T(\ell_1) > 27$, $40 \text{ GeV} < m_{\ell\ell} < 400 \text{ GeV} $ & 322 & 38 &  63 & 432\\
		$m_{jj} < 160$ GeV & 179 & 21 & 37 & 250\\
		$ee-mode$& 14 & 2 & 3 & 30\\
		$e\mu-mode$& 89 & 11 & 18 & 121\\
		$\mu\mu-mode$& 58 & 5 & 11 & 79\\
		\hline
	\end{tabular}
    \caption{Cutflow table $WWW$(ATLAS)~\cite{ATLAS:2022xnu} $2\ell$ SR.}
    \label{tab:ATLASWWW_2l}
\end{table*}

\begin{table*}[h!]
	\centering
	\begin{tabular}{l c c c c}
		\hline
		\textbf{Cut} & $\Delta^0 \Delta^\pm$ & $\Delta^0 \Delta^\pm$  & $\Delta^\mp
		\Delta^\pm$ & SM $WWW$ \\
        &  $WW$(100\%)  &  $WW$(0\%) &  & \\
		\hline
		$= 3$ leptons &  307 & 103 & 165 & 174\\
		$p_T(\ell_1) > 27$ & 306 & 103  & 164 & 174\\
		$b$-jet veto and no-OSSF leptons & 35  &2 & 5 & 38\\
		\hline
	\end{tabular}
    \caption{Cutflow table $WWW$(ATLAS)~\cite{ATLAS:2022xnu} $3\ell$ SR.}
    \label{tab:ATLASWWW_3l}
\end{table*}

\begin{table*}[h!]
	\centering
	\begin{tabular}{l c c c c c}
		\hline
		\textbf{Cut} & NP1 & NP2 & $WW(100\%)$  & $WW(0\%)$  & SM $tWZ$ \\
        & $pp \to t\bar{t}$ &  $pp \to\Delta^\pm\Delta^\mp$  & $pp \to\Delta^0\Delta^\mp$ & $pp \to\Delta^0\Delta^\mp$ & \\
        & $t\to\Delta^\pm b\,(0.01\%),$ & $\Delta^\pm\to tb$ & $\Delta^\pm\to tb,WZ$ & $\Delta^\pm\to tb,WZ$ & \\
        & $\bar{t}\to Wb$ &$\Delta^\mp\to WZ$ & $\Delta^0\to WW$ & $\Delta^0\to bb, ZZ$ &\\
		\hline
		at least 3 leptons & 80.1 & 309.56 & 580.9 & 291.69 & 198.91 \\
        $p_T(\ell_1,\ell_2) > (25,15)$ GeV &  78.53 & 303.63 & 568.02 & 285.48 & 197.85\\
        at least 1 OSSF with & 9.25 & 86.57 & 123.18 & 95.53 & 99.51 \\
        $|m_{\ell\ell} - m_Z| < 15$ GeV & &&  & & \\
        $= 3\ell$, $2j$, at least 1 $b$-jet & 3.66 & 10.51 & 7.96 &  16.3 & 41.59 \\
        $= 4\ell$, at least 1 $b$-jet & 0.87 & 1.87 & 1.89 & 3.67 & 7.0 \\
		\hline
	\end{tabular}
    \caption{Cutflow table $tWZ$(CMS)~\cite{CMS:2025xph} for Run~2.}
    \label{tab:CMStWZ_R2}
\end{table*}

\begin{table*}[h!]
	\centering
	\begin{tabular}{l c c c c c}
		\hline
		\textbf{Cut} & NP1 & NP2 & $WW(100\%)$  & $WW(0\%)$  & SM $tWZ$ \\
        & $pp \to t\bar{t}$ &  $pp \to\Delta^\pm\Delta^\mp$  & $pp \to\Delta^0\Delta^\mp$ & $pp \to\Delta^0\Delta^\mp$ & \\
        & $t\to\Delta^\pm b,\,(0.01\%)$ & $\Delta^\pm\to tb$ & $\Delta^\pm\to tb,WZ$ & $\Delta^\pm\to tb,WZ$ & \\
        & $\bar{t}\to Wb$ &$\Delta^\mp\to WZ$ & $\Delta^0\to WW$ & $\Delta^0\to bb, ZZ$ &\\
		\hline
		at least 3 leptons & 28.31  &  114.17  & 201.64 & 108.75 &  86.6\\
        $p_T(\ell_1,\ell_2) > (25,15)$ GeV & 28.07  & 113.82  & 200.84 & 108.35 & 86.6\\
        OSSF with & 3.53  & 33.28  & 43.34 &  37.1  & 43.57\\
        $|m_{\ell\ell} - m_Z| < 15$ GeV & &&  & & \\
        $= 3\ell$, $2j$, at least 1 $b$-jet &  1.33 & 4.19 & 3.72 & 6.5 & 18.43\\
        $= 4\ell$, at least 1 $b$-jet & 0.24  & 0.54 & 0.47 & 1.5 & 2.91 \\
		\hline
	\end{tabular}
    \caption{Cutflow table $tWZ$(CMS)~\cite{CMS:2025xph}  for Run~3.}
    \label{tab:CMStWZ_R3}
\end{table*}

\bibliographystyle{utphys}
\bibliography{EPJC_format}
\end{document}